\documentclass[useAMS,usegraphicx]{mn2e}
\newcommand{\beq}{\begin{equation}}
\newcommand{\beqa}{\begin{eqnarray}}
\newcommand{\eeq}{\end{equation}}
\newcommand{\eeqa}{\end{eqnarray}}

\newcommand{\apj}{ApJ}
\newcommand{\apjl}{ApJL}
\newcommand{\mnras}{MNRAS}
\begin{document}
%
\title [Effects of wind on radiation spectra from magnetized 
        accretion disks]
{Effects of wind on radiation spectra from magnetized 
        accretion disks}
\author[]{Naohiro Yamazaki$^{1,2}$, Osamu Kaburaki$^1$,
Motoki Kino$^{3}$\\
$^{1}$ Astronomical Institute, Graduate School of Science, 
Tohoku University, Aoba-ku, Sendai 980-8578, Japan\\
$2$ Present address: Iwaki-Shuei High School, Iwaki,
Fukushima 971-8185, Japan\\
$^{3}$ Department of Earth and Space Science,
Osaka University, Toyonaka 560-0043, Japan
}
\date{submitted to MNRAS 2002 June ??}
\twocolumn

\maketitle

\begin{abstract}
The effects of a wind on the emerging spectrum from an
inefficiently-radiating accretion flow in a global magnetic field
are examined, based on the analytic solution obtained recently by
one of the present authors. 
The results exhibit the steepening of the negative slope appearing 
in the intermediate frequency range of bremsstrahlung spectrum and 
the decrease in the luminosity ratio of thermal synchrotron to 
bremsstrahlung, in accordance with the increasing wind strength. 
Both effects are due to a suppressed mass accretion rate in the inner 
disk, caused by a mass loss in terms of wind. 

In order to demonstrate the reliability of this model, Sagittarius 
A$^*$ (Sgr A$^*$) and the nucleus of M\ 31, both of which have been 
resolved in an X-ray band by {\it Chandra}, are taken up as the best 
candidates for the broadband spectral fittings. 
Although the observed X-ray data are reproduced for these objects 
by both of the inverse-Compton and the bremsstrahlung fittings, 
some evidence of preference for the latter are recognized. 
The wind effects are clearly seen in the latter fitting case, in
which we can conclude that a widely extending accretion disk is
present in each nucleus, with no or only weak wind in Sgr A$^*$ and
with a considerably strong wind in the nuclear region of M\ 31. 
Especially in Sgr A$^*$, the inferred mass accretion rates are
much smaller than the Bondi rate, whose estimate has become reliable
due to {\it Chandra}.
This fact strongly suggests that the accretion in this object does
not proceed like Bondi's prediction, though its extent almost
reaches the Bondi radius. 

\end{abstract}


\section{INTRODUCTION}

The broadband spectrum of Sgr A$^*$, from radio to hard X-ray bands, 
was reproduced fairly well for the first time by Narayan, Yi and 
Mahadevan (1995) based on an optically thin ADAF (advection-dominated 
accretion flow) model. 
In contrast to the standard accretion-disk model (Shakura \& Sunyaev 
1973), the optically thin ADAF model could reproduce the wide spread 
of the observed spectrum. 
Namely, the rather narrow peak in the radio band was explained by 
thermal synchrotron radiation from the inner part of a disk with 
relativistic temperature, and X-ray luminosity, by the bremsstrahlung 
from the whole disk. 
Another great advantage of this model is in its low efficiency in 
producing radiative fluxes. 
The latter feature could make the model possible to explain the low 
luminosity of Sgr A$^*$ compared with that expected from a standard 
disk of about the Bondi accretion rate (see, e.g., Melia \& 
Falcke 2001). 

This type of models for optically thin ADAFs has been developed 
by many authors (e.g., Ichimaru 1977; Rees et al. 1982; Narayan \& 
Yi 1994, 1995a, b; Abramowicz et al. 1995; Blandford \& Begelman 
1999). 
Since, in this category, the viscosity of the accreting plasma plays 
a dominant role in both energy dissipation and angular-momentum 
extraction processes, we call it the viscous ADAF model. 
The model looked successful in explaining the broadband spectra not 
only of Sgr A$^*$ (Narayan, Yi \& Mahadevan 1995; Manmoto, Mineshige 
\& Kusunose 1997; and Narayan et al. 1998b), but also of Galactic 
X-ray sources and of low-luminosity active galactic nuclei (LLAGNs; 
for a review, see Narayan, Mahadevan \& Quataert 1998a). 

Meanwhile, if the presence of ordered magnetic fields in the nuclear 
regions of galaxies is taken seriously, anther type of optically 
thin ADAF model can be constructed (Kaburaki 2000, for brevity, K00). 
Since, in this model, the gravitational energy of accreting matter 
is liberated through the plasma's electric resistivity it will be 
called the resistive ADAF model. 
Angular momentum, on the other hand, is extracted from the accreting 
matter by an ordered magnetic field that is penetrating the disk and 
is twisted, to a certain extent, by the rotational motion of the 
accreting matter. 
The observed spectrum of Sgr A$^*$ could be reproduced also by this 
model (Kino, Kaburaki \& Yamazaki 2000, for brevity, KKY) as 
satisfactorily as the viscous ADAF models could. 

Recently, however, a fairly drastic change of the above stated situation 
has come with the appearance of the results of the X-ray telescope, 
{\it Chandra}. 
Owing to its high resolution, the apparent nuclear sources of a few 
nearby galaxies have been further resolved into several point sources 
including a true nuclear source in each case (e.g., Garcia et al. 2000 
for M31; Baganoff et al. 2001a for Sgr A$^*$).
The intrinsic X-ray luminosity of these true nuclear sources are 
therefore considerably lower in fact, and moreover, their spectra 
have tuned out to be much softer than previously believed. 

Especially, this softness of the spectra requires a critical 
reconsideration of the broadband spectral fittings of these objects 
by optically thin ADAF models, because this fact may exclude the hitherto 
accepted bremsstrahlung fitting to the X-ray band of the spectra. 
Further, as already been demonstrated by some authors (e.g., Quataert 
\& Narayan 1999; Di Matteo et al. 2000), the presence of winds emanating 
from accretion disks can alter the ratio of X-ray luminosity to radio 
luminosity. 
Therefore, it is also necessary to include the presence of winds into 
the basic models based on which the spectral fittings are performed. 
In this context, the resistive ADAF model has been developed to 
include winds emanating from the disk surfaces (Kaburaki 2001, for 
brevity, K01).

The present paper is devoted to describe the general predictions on 
the broadband spectra radiated from such accretion disks as described 
by the analytic model of K01, and to report the results of applications 
to Sgr A$^*$ and the nucleus of M31. 
Although Di Matteo et al. (2000) insist the presence of winds in the 
nuclei of some nearby elliptical galaxies on the basis of their 
broadband spectral fittings by a wind-version of the viscous ADAF model, 
the results are still uncertain because the X-ray fluxes they used are 
obtained by {\it ASCA} and hence the nuclear sources are not resolved 
(see also Quataert \& Narayan 1999, for a Galactic X-ray transient and 
Sgr A$^*$). 
We therefore restrict our applications only to such objects as have 
been resolved by {\it Chandra} and the results of observations have been 
open to the public.

\section{RESISTIVE ADAF MODEL INCLUDING WINDS}

In this section, we first introduce the basic ideas and physics 
contained in the resistive ADAF model proposed in K01, in which 
the presence of winds from the disk surfaces is allowed for. 
A schematic drawing of the global configuration presumed in the this 
model is given in Figure 1. 
An asymptotically uniform magnetic field is vertically penetrating 
the accretion disk and is twisted by the rotational motion of the plasma. 
Owing to the Maxwell stress of this twisted magnetic field, a certain 
fraction of the angular momentum of accreting plasma is carried out 
to infinity, and this fact ensures the plasma to gradually infall toward 
the central black hole. 

In a stationary state, the deformation of magnetic field lines is 
determined by a balance between motional dragging and diffusive 
slippage of the field lines. 
Since the representative magnetic-Reynolds number $\Re$ is large 
(i.e., $\Re^2(r) \gg 1$) in the disk (except the region close to its 
inner edge, where $\Re \sim 1$), the deformations are also large. 
In this sense, the disk can be said as weakly resistive. 
Generally, the deformation in the toroidal direction is larger than 
that in the poloidal direction (i.e., $b_{\varphi}/b_{\rm p}\sim\Re$, 
where $b$ denotes the deformed part of the magnetic field). 

The vertical structure of the disk is maintained by a pressure balance 
between the magnetic pressure of the toroidal field, which is dominant 
on the outside the disk, and the gas pressure in the disk: i.e., the 
accreting plasma is magnetically confined in a geometrically thin disk. 
Reflecting this fact, the gas pressure and density in the disk become 
quantities of order $\Re^2$ (see, equations [49] and [58] in K01). 
In general, this balance is not a static balance in its strict sense, 
and there may be a vertical flow from the upper and lower surfaces of 
a disk. We call such outflows (or inflows depending on the case) winds, 
distinguished from jets that may be formed within the inner edge of 
the accretion disk (also see below). 

Since the wind velocity obtained in K01 is much smaller than the 
rotational velocity (by a factor of order $\sim\Delta\Re^{-1}$, 
where $\Delta\ll1$ is the half-opening angle of the disk), its inertial 
force can hardly affect the vertical force balance described above. 
However, the wind may be accelerated, by some mechanisms with which 
we do not concern in this paper, to a considerable speed after it has 
been injected from the accretion disk to the wind region outside 
the disk. 
Here, we only expect that the upward wind proceeds to infinity because 
its total energy per unit mass (i.e., the Bernoulli sum in K01) is 
positive and hence the flow is unbound in the gravitational field. 

The rotational velocity is a certain fraction of the Kepler velocity 
(i.e., a reduced Keplerian rotation) because the radial pressure-gradient 
force, together with the centrifugal force, sustains the gravitational 
pull on the plasma. 
In contrast to the rotational velocity, the infall velocity is a small 
quantity of the order of $\Re^{-1}$ as far as the disk is weakly 
resistive (i.e., except for the region near the inner edge). 
Their ratio does not depend on the disk thickness $\Delta$ as far as 
$\Re$ is regarded as a free parameter like the viscosity parameter 
$\alpha$ in the viscous ADAF models. 

It has been demonstrated in K01 that, if the flow is completely adiabatic 
(i.e., if the liberated gravitational energy remains in each fluid element 
and is merely advected down the flow), the flow cannot drive winds. 
On the other hand, if some mechanism of energy transport (i.e., 
non-adiabaticity) allows the energy flow toward the accretion disk from 
the region within its inner edge, the disk can drive an upward wind.
In K01, the presence of such mechanisms is treated implicitly 
in terms of a parameter $n$ that specifies the strength of a wind. 
The essence of the results is in that the presence of such 
non-adiabaticity affects solely on the radial profiles of the quantities 
such as density, pressure and magnetic field components (velocity 
components and temperature are not affected) but not on the vertical 
profiles. 
Thus, the presence of a wind appears unambiguously as a radius dependent 
mass accretion rate. 

The analytic solution obtained in K01 describes the physical quantities 
in the accretion disk (i.e., the solution is valid only in between 
the inner and outer edges of the disk) with also a decreasing accuracy 
toward the upper and lower surfaces of the disk. 
The latter part of this statement means that the solution admits 
a few inconsistencies of the order of $\tanh^2\eta$, where $\eta\equiv 
(\theta-\pi/2)/\Delta$, in its variable-separated forms. These are 
negligible, however, except near the disk surfaces. 
Further, the solution does not care about the radiation loss. 
Nevertheless, since it has been confirmed retrospectively that the expected 
radiation flux from such a disk is negligibly small as far as the accretion 
rate is sufficiently smaller than the Eddington rate, the solution is 
consistent within this restriction. 
The disk spectrum can, therefore, be calculated with sufficient accuracy 
by using this solution, without any correction to the radiation losses. 

If we extrapolate our physical understanding obtained within the disk 
even to the surrounding space, we are naturally led to the following 
picture of a galactic central engine. 
The engine is essentially a hydroelectric power station in which 
the ultimate energy source is the potential energy of the accreting 
plasma in the gravitational field. 
The accretion disk is a DC dynamo of an MHD type and drives a poloidally 
circulating current system, which is the cause of the toroidal magnetic 
component added to the originally vertical field (i.e., the twisting 
of the field lines). 
In the configuration shown in Figure 1, the radial current is driven 
outward in the accretion disk, and a part of the current closes its 
circuit through the near wind region while another part closes after 
circulating remote regions (probably reaching the boundary of a
`cocoon' enclosing hot winds). 
Anyway, these return currents concentrate within the polar regions in the 
upper and lower hemispheres, and finally return to the inner edge of 
the disk. 

A bipolar jet may be formed from the plasma in this polar current regions, 
because the Lorentz force due to the toroidal field always has both 
of the necessary components for collimation and for radial acceleration, 
as shown in the figure. 
The often suggested universality of the association of an accretion 
disk and a bipolar jet is thus understood naturally in terms of one 
physical entity, the {\it poloidally circulating current system}. 
It is very likely that only a small fraction of the infalling 
matter input in the disk can actually fall onto the central black hole, 
with the remaining part expelled as a bipolar jet and a wind from 
disk surfaces.

\section{SCALED QUANTITIES}

The radial and co-latitude dependences of the all relevant quantities 
in the resistive ADAF solution are given in K01. 
In rewriting these quantities in a scaled form convenient for the 
discussion of LLAGNs, the following normalizations are adopted: 
\begin{equation}
  x \equiv \frac{r}{r_{\rm S}}, \quad 
    m\equiv \frac{M}{10^{8}M_{\odot}}, \quad
     \dot m\equiv \frac{\dot{M}_0}{\dot{M}_{\rm c}}, \quad
       b \equiv \frac{\vert B_{0}\vert}{1G}, 
\end{equation}
where $r$, $M$, $\dot{M}_0$ and $B_0$ denote the radial distance, 
black-hole mass, mass accretion rate at the disk's outer edge, 
$r=r_{\rm out}$, and strength of the external magnetic field at 
$r_{\rm out}$, respectively. 
The radius is normalized by the gravitational radius of the central 
black hole (actually neglecting its rotation, by the Schwarzschild 
radius $r_{\rm S}$) and the accretion rate, by the critical accretion 
rate defined in terms of the Eddington luminosity $L_{\rm E}$ as 
$\dot{M}_{\rm c} \equiv L_{\rm E}/( 0.1 c^{2})$, taking into account 
a typical conversion efficiency of $0.1$. 

Further, we eliminate $\vert B_0\vert$ by the aid of the relation (see K01)
\begin{equation}
  r_{\rm out} = \left( \frac{3\Re_0^{4n}}{2n+1}\ 
   \frac{GM\dot{M}_0^2}{B_0^4}\right)^{1/5}, 
\end{equation}
or 
\begin{equation}
  \quad b = 7.6\times10^4\ (2n+1)^{-1/4}\Re_0^n
   x_{\rm out}^{-5/4}\dot{m}^{1/2}m^{-1/2}, 
\end{equation}
where $G$ is the gravitational constant. 
This is because $B_0$ seems rather difficult to infer from observations 
compared with the outer edge radius $r_{\rm out}$. 
Substituting the latter expression into the radial part of the 
variable-separated forms of the relevant quantities, we finally obtain 
\begin{equation}
   \tilde{v}_{\varphi}(x)= 1.2\times 10^{10}\ (2n+1)^{1/2} x^{-1/2} 
     \quad{\rm cm}\ {\rm s}^{-1},
\end{equation}
\begin{eqnarray}
   \tilde{\rho}(x)&=& 1.5\times 10^{-12}\ (2n+1)^{-1/2}
   \Re_0^{2(n+1)} \nonumber \\
 &&\times x_{\rm out}^{-(1/2+2n)}\dot{m}m^{-1}x^{-(1-2n)} 
   \quad{\rm g}\ {\rm cm^{-3}},
 \label{eqn:dens}
\end{eqnarray}
\begin{eqnarray}
   \tilde{p}(x)&=& 2.3\times10^8\ (2n+1)^{-1/2}\Re_0^{2(n+1)}\nonumber \\
   &&\times 
x_{\rm out}^{-(1/2+2n)}\dot{m}m^{-1}x^{-2(1-n)} \quad{\rm dyne\ cm}^{-2},
\end{eqnarray}
\begin{equation}
   \tilde{T}(x)= 9.0\times10^{11} x^{-1} \quad{\rm K},
  \label{eqn:T}
\end{equation}
\begin{eqnarray}
   \vert \tilde{b}_{\varphi}(x)\vert &=& 7.6\times10^4\ (2n+1)^{-1/4} 
   \Re_0^{n+1}\nonumber \\
&&\times x_{\rm out}^{-(1/4+n)}\dot{m}^{1/2}m^{-1/2}\ x^{-(1-n)} 
   \quad{\rm G},
  \label{eqn:B}
\end{eqnarray}
and further, 
\begin{equation}
   \tau_{\rm es} (x) = 3.7\times10^1\ (2n+1)^{-1/2}\Re_0 
   x_{\rm out}^{-(1/2+2n)}\dot{m}x^{2n}, 
\end{equation}
\begin{equation}
   \dot{M}(x) = \dot{M}_0x_{\rm out}^{-2n}x^{2n}, 
 \label{eqn:Mdot}
\end{equation}
\begin{equation}
   \Delta = \Re_0^{-(2n+1)}, 
\end{equation}
where $v_{\varphi}$, $\rho$, $p$ and $\Delta$ are the rotational 
velocity, density, gas pressure and the half-opening angle of the disk, 
respectively, and the tildes above them denote their radial parts. 
The opacity is dominated by the electron scattering, and the resulting 
optical depth in the vertical direction $\tau_{\rm es}(x)$ depends on 
the radius like $x_{\rm out}^{-2n}x^{2n}=(r/r_{\rm out})^{2n}$. 
Therefore, it is much smaller than unity as far as $\dot{m}\ll 1$ 
and $x_{\rm out}\gg 1$ since $r/r_{\rm out} < 1$. 
We can confirm the explicit dependence of the accretion rate on the 
radius in equation (\ref{eqn:Mdot}). 

Since the radii of inner and outer edges are mutually related by 
$x_{\rm in} = \Re_0^{-2} x_{\rm out}$ (see K01), they are written in the 
present units as 
\begin{eqnarray}
  r_{\rm out}&=& 3.0\times10^{13}\ x_{\rm out} m \quad\mbox{cm}, \nonumber \\
  \qquad r_{\rm in}&=& 3.0\times10^{13}\ \Re_0^{-2}x_{\rm out} m 
  \quad\mbox{cm}. 
\end{eqnarray}
Although the numerical values appearing in the above scaled expressions 
are their formal values at the black hole horizon, a direct 
inward-extension of the above solution beyond the inner edge is
meaningless because the solution becomes invalid there.  
Finally, we note that the number of free parameters contained in our 
solution is altogether 5, i.e., $m$, $\dot{m}$, $x_{\rm out}$, $\Re_0$ 
and $n$. 

\section{RADIATION PROCESSES}

The method of calculating the expected spectrum from an accretion 
disk, based on the analytic solution of the resistive ADAF model 
including winds, is essentially the same as stated in KKY for the 
no-wind version of that model. 
The main radiation mechanisms are thermal cyclo-synchrotron emission 
most of which contributes to the flux in the radio band, thermal 
bremsstrahlung that may contribute mainly in the X-ray band, and 
inverse Compton scattering of the synchrotron (and bremsung) photons,
which may appear as one or two peaks in the middle frequency range 
(and the scattered bremsung flux that may be scarcely recognized 
above the exponentially decaying part of the spectrum). 

There are in the present scheme, however, two main differences from 
the former version. 
One is an improvement in the treatment of the Gaunt factor, 
and the other is in the approximation used in evaluating the 
unscattered (the sum of synchrotron and bremsung processes) flux 
from a disk.  
They are described in the following subsections. 
Hereafter, we need only the radial parts of the relevant physical 
quantities, and hence quote them without tildes on them for simplicity. 

We first assume that an observed spectrum from a galactic nucleus is 
produced only by its accretion disk, and neglect the possible 
contributions from other components such as jet or wind. 
The validity of this assumption will be discussed in \S5.3. 

\subsection{Unscattered Flux}

Since the Compton unscattered flux consists of the synchrotron emission 
and the bremsung (both of which are assumed to be isotropic), the 
the absorption coefficient is calculated from Kirchhoff's law 
\begin{eqnarray}
  \alpha_{\nu} = \frac{\chi_{\nu, {\rm br}}+\chi_{\nu, {\rm sy}}}
    {4\pi B_{\nu}},
\end{eqnarray}
where $\chi_{\nu}$ denotes the volume emissivity for both processes and 
$B_{\nu}$ is the Planck function. 
Assuming LTE, we have for the specific intensity 
\begin{eqnarray}
  I_{\nu} = B_{\nu}\left[1-\exp{(-\tau_{\nu})}\right],
\end{eqnarray}
where $\tau_{\nu}=\tau_{\nu}^*/\cos\theta$, $\tau_{\nu}^*=\alpha_{\nu}H$ 
and $H$ is the height of the disk. 

In this situation, we approximate the emerging flux by an interpolation 
formula as 
\begin{eqnarray}
  F_{\nu} \simeq \pi B_{\nu}\left[1-\exp{(-2\tau_{\nu}^*)}\right]. 
\end{eqnarray}
This formula has a merit of giving the correct answers $\pi B_{\nu}$ 
and $(\chi_{\nu,{\rm br}}+\chi_{\nu,{\rm sy}})H/2$ in the optically 
thick and thin limits, respectively. 
The difference between this and the two-stream approximation adopted 
in KKY, however, remains small.

\subsection{Bremsstrahlung}

According to Narayan \& Yi (1995b) and Manmoto, Mineshige \& Kusunose 
(1997), KKY used the non-relativistic version of the Gaunt factor 
$g_{\rm ff}(\nu, T)$ not only for electron-ion collisions but also 
for electron-electron collisions, and a constant frequency-integrated 
Gaunt factor $g_{\rm B}(T)$. 
Compared with the more strict treatments of these quantities by 
Skibo et al. (1995), the old scheme evidently over-estimates the 
contribution from the e-e process when $h\nu<m_{\rm e}c^2$ (where 
$h$ is the Planck constant and $m_{\rm e}$ is the electron mass), 
and under-estimates it by about one order of magnitude when
$h\nu>m_{\rm e}c^2$ (for more details, see Yamazaki 2002).
Therefore, we adopt in this paper the formulae given by Skibo et al. 
(1995).
The effect of this change on a spectrum is illustrated in Figure 7 
(the thick and thin solid curves are calculated by the new and old
versions, respectively, for the same fitting parameters). 

In order to grasp the basic ideas, we first assume that the Gaunt 
factor is nearly constant even in the relativistic temperature regime. 
Then, the luminosity due to bremsstrahlung is roughly written as 
\begin{equation}
  L_{\nu}^{\rm bs} \propto \int_{r_{\rm in}}^{r_{\rm out}} 
    \rho^2 T^{-1/2}\exp[-h\nu/k_{\rm B}T]\ dV 
  \propto f\ \Re_0^{2n+3}\dot{m}^2 m, 
\end{equation}
where $k_{\rm B}$ is the Boltzmann constant, $dV=4\pi\Delta r^2dr$ 
is the volume element in the disk, and 
\begin{equation}
  f \equiv \int_{x_{\rm in}}^{x_{\rm out}}x^{1/2+4n}
    \exp[-h\nu/k_{\rm B}T(x)]\ dx. 
\end{equation}
Properly evaluating this integral $f$ in each of the typical frequency 
ranges, we have the frequency dependence 
\begin{equation}
  f \propto \cases{
  \mbox{const.}:& $\qquad \nu \ll k_{\rm B}T_{\rm out}/h$  \cr
  \nu^{-3/2-4n}:& $\qquad 
    k_{\rm B}T_{\rm out}/h < \nu \ll k_{\rm B}T_{\rm in}/h$ \cr
  \exp[-h\nu/k_{\rm B}T_{\rm in}]:& 
    $\qquad \nu > k_{\rm B}T_{\rm in}/h.$ \cr}
 \label{eqn:I}
\end{equation}
In fact, however, in addition to the above features, there may appear an 
increase in the luminosity in the frequency range $\nu \sim 
k_{\rm B}T_{\rm in}/h$, which is due to the increase of the Gaunt factor 
at relativistic temperatures (see the peak B$^\prime$ in Figure 3). 

The predicted power-law index of the luminosity $\nu L_{\nu}^{\rm bs}$, 
i.e. $-1/2-4n$ on the frequency, in the intermediate frequency range 
($k_{\rm B}T_{\rm out}/h < \nu \ll k_{\rm B}T_{\rm in}/h$) exhibits an 
explicit difference from the corresponding prediction by the viscous 
ADAF model including winds (Quataert \& Narayan 1999), $1/2-2p$ (where 
$p=2n$). 
In the no-wind case ($n=p=0$), the former predicts a negative slope 
while the latter does a positive slope in the $\nu$-$\nu L_{\nu}$ 
diagram. 
This difference comes from the difference in the radial dependences 
of the density in the basic analytic solutions (K00 and Narayan \& Yi 
1994, respectively). 
In both models, the inclusion of a wind ($n>0$) equally reduces these 
values depending on its strength. 
This is because the presence of a wind reduces the contribution from 
the inner parts of a disk to the bremsstrahlung (since $\dot{M}\propto 
x^{2n}$), resulting in a reduction of the higher frequency side of 
the spectrum. 

\subsection{Synchrotron Emission}

The adopted formulae for the calculation of synchrotron emission 
and inverse Compton scattering is the same as in KKY. 
This synchrotron formula is known to be accurate when the electron 
temperature $T_{\rm e}$ is in the range $10^8<T_{\rm e}<3\times10^{10}$. 
This temperature range corresponds in our solution to the radius 
range of $30<x<10^4$. 
Since, however, the contribution from the lower temperature region 
($x>10^4$) is negligibly small compared with that from the bremsung, 
we can extend our calculation even to such regions (until the effects 
of recombination on the bremsung becomes important), by safely truncating 
the cyclo-synchrotron emission. 

The critical frequency under which the radiation becomes optically 
thick at a fixed $r$ is given (Narayan \& Yi 1995b; Mahadevan 1997) by 
\begin{eqnarray}
 \nu_{\rm c} \propto T^2(x)\vert B_{\varphi}(x)\vert 
  \propto \Re_0^{n+1}x_{\rm out}^{-(1+4n)}\dot{m}^{1/2}m^{-1/2}x^{-(3-n)}. 
\end{eqnarray}
Since $\nu_{\rm c}$ increases towards the inner edge, inner part of 
a disk is apt to be optically thick at a given frequency $\nu$, 
with a critical radius 
\begin{eqnarray}
  x_{\rm c} &\propto& \Re_0^{(n+1)/(3-n)}
  x_{\rm out}^{-(1+4n)/(3-n)}\nonumber \\
   && \times \dot{m}^{1/2(3-n)}m^{-1/2(3-n)}\nu^{-1/(3-n)}.
 \label{eqn:xc}
\end{eqnarray} 
Anyway, there is a radius $x_{\rm p}$ near $x_{\rm in}$ whose 
contribution to the self-absorbed radiation is the largest so that the 
peak frequency is determined as $\nu_{\rm p}=\nu_{\rm c}(x_{\rm p})$. 

Then, for the self-absorbed luminosity per frequency ($\nu_{\rm min}<\nu
<\nu_{\rm p}$, where $\nu_{\rm min}=\nu_{\rm c}(x_{\rm out}$)), 
we have 
\begin{equation}
  L_{\nu}^{\rm abs} \propto 
  \int_{r_{\rm in}}^{r_{\rm c}}\frac{\nu^2}{c^2}k_{\rm B}T\ 2\pi 
  r\ dr \propto (x_{\rm c}-x_{\rm in})m^2\nu^2 
  \approx x_{\rm c}m^2\nu^2,  
 \label{eqn:slabs}
\end{equation}
where the final expression is valid only in the frequency range 
$\nu\ll\nu_{\rm p}$ in which $x_{\rm c}\gg x_{\rm in}$. 
In the crude approximation in which $x_{\rm c}$ is regarded as a constant, 
the luminosity of the self-absorbed synchrotron emission is determined 
solely by the central mass $m$. 

However, substituting the above estimation for $x_{\rm c}$ in equation 
(\ref{eqn:xc}), we have a more accurate dependence 
\begin{eqnarray}
  L_{\nu}^{\rm abs} &\propto& 
  \Re_0^{(n+1)/(3-n)}x_{\rm out}^{-(1+4n)/(3-n)} \nonumber \\ 
&& \times \dot{m}^{1/2(3-n)}
   m^{(11-4n)/2(3-n)}\nu^{(5-2n)/(3-n)}.  
 \label{eqn:mslabs}
\end{eqnarray}
The strongest dependence on the parameters is still on the central mass, 
and its power varies slightly from 1.80 to 1.85 for the variation of $n$ 
in the allowed range, from 1/2 to $-1/4$. 
The dependence on the other parameters are rather weak, and this fact 
is very favorable to determine, at least within the framework of the 
present model, the central mass from the spectral fittings of this 
self-absorbed part, without much ambiguity (however, see also \S
5.3). 
The power of $\nu$ also varies only slightly from 1.6 to 1.7 in the same 
range of $n$. 

The main effect of a wind on the synchrotron spectrum is a reduction 
of its peak intensity (when $n>0$) according to the wind strength. 
This is because the synchrotron emission comes mainly from the inner 
most region of an accretion disk where the magnetic field is the 
strongest and the electrons are most energetic, so that the reduction 
of the accretion rate in this region due to wind loss causes a decrease 
in the optically thin part of the emission.

\section{APPLICATIONS}

As stated in the previous section, we can calculate the spectrum
emanating from an accretion disk by specifying the values of
five parameters, $m$, $\dot{m}$, $x_{\rm out}$, $\Re_0$, and
$n$.
In other words, we can determine these values from the process of
spectral fitting only if we have enough data points distributed over 
the full range of the spectrum of a specific object.
In the applications of our model to Sgr A$^*$ and M\ 31, we first
proceed with this spirit, and regard all five parameters as free
fitting parameters.

In fact, however, there are other information independent of
the spectral data.
For example, the central black-hole masses are known for these
objects fairly accurately by the observations based on the dynamics.
In the case of Sgr A$^*$, the {\it Chandra} observations provide
us a  good estimate of the Bondi accretion rate.
In principle, the results obtained from a spectral fitting should
not conflict with these additional informations.
If, however, there are some discrepancies between them, it may
offer some important information about the inadequacy of the
present status of the basic model or of our general understanding.
Such considerations will be given in \S 5.3. 

\subsection{Sgr A$^*$}

{\it Chandra} has resolved a weak source at the radio position of 
Sgr A$^*$ within the accuracy of 0$^{\prime\prime}$.35 (Baganoff 
et al. 2001a). 
Its absorption-corrected luminosity in 2-10 keV band is 
$2.4_{-0.6}^{+3.0}\times 10^{33}$ ergs s$^{-1}$ and power-law fitted 
photon index is $2.7_{-0.9}^{+1.3}$. 
For the other observational data than this X-ray band, we use those 
compiled by Narayan et al. (1998b). 
For the sake of comparison, the {\it ROSAT} value is also plotted 
in Figures 2, 3 and 7 with an empty circle. 
Among the radio observations, the 86 GHz point with the highest 
resolution obtained by VLBI is taken most seriously in the following 
spectral fittings. 

The most restrictive feature of the {\it Chandra} results to the 
spectral fittings is the softness of the X-ray spectrum whose most 
probable slope in the $\nu$-$\nu L_{\nu}$ diagram is negative, in 
contrast to the {\it ROSAT} observation in which the slope in a similar 
X-ray range is positive. 
Although a rapid X-ray flaring is reported (Baganoff et al. 2001b), 
we do not discuss it here because the event is a transient phenomenon 
localized in a small region of the accreting plasma. 

One possible way of reproducing this negative slope by our model is to 
use the decreasing side of the first-order Compton peak. 
Hereafter, we call this type of fittings the `Compton fitting'. 
In order for this negative slope to reach the {\it Chandra} X-ray band, 
the optically thin (i.e., the high frequency) side of the synchrotron 
peak should be located in a high enough frequency range. 
Nevertheless, the location of its self-absorbed (i.e., the low frequency) 
side is fixed by the 86 GHz point. 
Thus, the synchrotron peak in this fitting is required to be wide enough. 
Also, in order to keep the required X-ray luminosity, we require a 
sufficiently high luminosity to the synchrotron peak. 
However, these two requirements apt to contradict with the upper limit 
data in IR band, and hence the Compton fitting becomes very tight. 
We do not include wind in this type of fittings, because its inclusion 
does not cause any change in the overall spectral shape but reduces
the luminosity only similarly at every frequency.  

Actually the above stated requirement allows only a unique fitting, 
which is shown in Figure 2 with the full curve.
The contribution from bremsstrahlung is buried in the second Compton
peak. 
This curve seems to be marginally fitted to  the observational
constraints. 
The curve traces the observational data points fairly well on the 
low frequency side of the radio peak except around its top, but the 
X-ray slope is nearly at its steepest limit. 
If we try to improve the fitting to the X-ray slope, it becomes
impossible for the curve to go through the 86  GHz point.

In this fitting, the accretion rate $\dot{m}$ has to be reduced from 
the previously inferred value (KKY) to a considerably lower value 
in order to suppress the contribution from bremsstrahlung. 
The position of the low frequency side of the synchrotron peak is fixed 
by adjusting mainly the black hole mass $m$ (see Eq. [\ref{eqn:slabs}] 
or [\ref{eqn:mslabs}]). 
The most effective parameter to shift the Compton peak-frequency is 
$x_{\rm out}$, and a shift toward the high frequency side means a 
reduction of $x_{\rm out}$.   
In order to avoid the appearance of the inner edge radii smaller than 
that of the marginally-stable circular orbit, the magnetic Reynolds
number at the outer boundary, $\Re_0$, should stay at rather small
values. 
Since the obtained ratio $x_{\rm out}/x_{\rm in}$ from this fitting 
is not so large, the amplification of the seed magnetic field $B_0$ 
by the sweeping and twisting effects of the accretion flow remains to be 
rather small. 
In other words, this case of Compton fitting requires considerably large 
external magnetic fields, since $b_{\varphi}(x_{\rm in})$ is almost fixed 
by the height of the synchrotron peak. 

The resulting values of the fitting parameters are summarized in
Table 1.
When a figure is expressed as three-digit number, it means that a change 
in the final digit causes a noticeable shift of the fitting curve. 
The dimension-recovered values and related physical quantities of
our interest are 
\begin{eqnarray}
  M = 6.0\times10^5\ M_{\odot}, \quad 
  \dot{M}_0 = 9.6\times10^{-9}\ M_{\odot}\ \mbox{yr}^{-1}, \nonumber\\
  r_{\rm out} = 1.8\times10^{13}\ \mbox{cm}, \quad 
  r_{\rm in} = 1.2\times10^{12}\ \mbox{cm} = 6.8\ r_{\rm S}, \nonumber\\
  \vert B_0\vert = 2.7\ \mbox{G}, \quad 
  \vert b_{\varphi}(r_{\rm in})\vert = 1.5\times10^{2}\ \mbox{G}, \nonumber\\
  \Delta = 2.60\times10^{-1}\ \mbox{rad}.
\end{eqnarray}

Another type of possible fittings is called the `bremsstrahlung fitting', 
in which the {\it Chandra} spectrum is reproduced by the intermediate 
frequency range of the bremsstrahlung (Eq. [\ref{eqn:I}]). 
In order for the {\it Chandra} frequency band to fall in this negative 
slope region, we have to take the disk's outer edge $x_{\rm out}$ so 
large that its temperature decreases below about 1 keV. 
This corresponds to the outer edges of $x_{\rm out} > 10^5$, and these 
values make a great contrast with the former value of $2.6\times10^3$, 
that was obtained in KKY in reproducing the positive X-ray slope of 
ROSAT result by the low frequency range of the bremsstrahlung. 
The relatively faint X-ray luminosity of {\it Chandra} means a fairly 
small $\dot{m}$ also in this case, but not so extremely as in the 
Compton fitting case. 

Two good examples of such fittings are shown in Figure 3. 
The full curve is for a no-wind case, and the dashed curve is for 
a weak-wind case. 
Although the fitting to the X-ray slope is much improved by the 
inclusion of a weak wind, the fitting to the low frequency side of 
the synchrotron peak becomes rather worse around its top. 
At present, we cannot clearly point out which is the best fit to 
Sgr A$^*$, because of the lack of data points that are effective 
to distinguish them. 
High resolution observations in sub-millimeter to IR band would be 
very useful not only for this purpose but also to decide the 
superiority between the Compton and bremsung fittings. 

The values of the fitting parameters obtained for each case are
cited in Table 1, and the resulting values for the physical
quantities are 
\begin{eqnarray}
  M = 4.0\times10^5\ M_{\odot}, \quad 
  \dot{M}_0 = 3.7\times10^{-8}\ M_{\odot}\ \mbox{yr}^{-1}, \nonumber\\
  r_{\rm out} = 1.2\times10^{16}\ \mbox{cm}, \quad 
  r_{\rm in} = 3.4\times10^{12}\ \mbox{cm} 
    = 2.9\times10^1\ r_{\rm S}, \nonumber\\
  \vert B_0\vert = 1.4\times10^{-3}\ \mbox{G}, \quad 
  \vert b_{\varphi}(r_{\rm in})\vert = 2.8\times10^{2}\ \mbox{G}, \nonumber\\
  \Delta = 1.70\times10^{-2}\ \mbox{rad}, 
\end{eqnarray}
for the solid curve, and 
\begin{eqnarray}
  M = 4.0\times10^5\ M_{\odot}, \quad 
  \dot{M}_0 = 2.0\times10^{-8}\ M_{\odot}\ \mbox{yr}^{-1}, \nonumber\\
  r_{\rm out} = 1.2\times10^{16}\ \mbox{cm}, \quad 
  r_{\rm in} = 1.6\times10^{12}\ \mbox{cm} 
    = 1.3\times10^1\ r_{\rm S}, \nonumber\\
  \vert B_0\vert = 1.6\times10^{-3}\ \mbox{G}, \quad 
  \vert b_{\varphi}(r_{\rm in})\vert = 4.2\times10^{2}\ \mbox{G}, \nonumber\\
  \Delta = 4.71\times10^{-3}\ \mbox{rad}, 
\end{eqnarray}
for the dashed curve. 

In order to maintain a relatively higher synchrotron luminosity compared 
with the X-ray luminosity in the spectrum of Sgr A$^*$, considerably  
large values are required for $\Re_0$ (i.e., small values for $\Delta$),
and this fact results in reasonably small values for $x_{\rm in}$
in spite of large $x_{\rm out}$. 
The latter fact (i.e., $x_{\rm in} \ll x_{\rm out}$), in turn,
guarantees the appearance of a well extended negative-slope region in
the spectrum, and comfortably small values for the external magnetic
field compared with the case of Compton fitting.  
The total bremsstrahlung hump in such a case shows a double-peaked
structure in which the second peak (B$^{\prime}$ in Figure 3) on the
higher-frequency side is emphasized by the enhancement of the Gaunt
factor in the relativistic temperature regions located in the inner
part of the disk (see also Figure 7). 

The results in this case, therefore, predict a fairly wide extent of 
the ADAF state up to $x_{\rm out} > 10^5$ with also a fairly low
accretion rate of the order of $\dot{m}\sim$ a few $\times10^{-6}$. 
For the obtained black-hole mass of $\sim$ a few $\times 10^5 M_{\odot}$, 
the former value corresponds to the size of $r_{\rm out} > 10^{16}$ cm, 
which is in good agreement with the marginally resolved radius of 
0.02 pc of the nuclear X-ray source (Baganoff et al. 2001a). 
The corresponding values of the mass accretion rate at the outer 
radius, $\dot{M}_0$, is $\sim$ a few $\times 10^{-8}M_{\odot}$yr$^{-1}$. 
In the weak wind case, this value is further reduced to $\dot{M}
(r_{\rm in})=3.4\times10^{-9}M_{\odot}$yr$^{-1}$ at the inner edge. 
Therefore, these values are consistent with the limit $\leq 10^{-8}
M_{\odot}$yr$^{-1}$ recently imposed on $\dot{M}(r_{\rm in})$ 
(Agol 2000; Quataert \& Gruzinov 2000) from the interpretation of 
polarization measurement in the radio emission (Aitken et al. 2000,
however see also Bower et al. 2001). 

\subsection{M\ 31}

The nuclear source observed by {\it ROSAT} at the center of M\ 31 
has also been resolved into five point sources by {\it Chandra} 
(Garcia et al. 2000). 
The true nuclear source is identified with that located within 
1$^{\prime\prime}$ of the supermassive black hole and has an 
anomalously soft spectrum. 
They report that its luminosity in the 0.3-7.0 keV band is $4.0_{-2.8}
^{+12}\times10^{37}$ erg s$^{-1}$ and the power-law fitted photon index 
is $4.5\pm1.5$ in the 0.3-1.8 keV range (we assume this value to hold 
up to 7.0 keV), after corrected for interstellar absorption. 
Although there seems to be some debates about these results, we 
tentatively fit our model to this observation. 

Other observational data are obtained by {\it VLA} (Crane et al. 1992, 
1993), {\it IRAS} (Soifer et al. 1986; Neugebauer et al. 1984), {\it KPNO} 
(McQuade, Calzelli \& Kinney 1995), Palomar 5m telescope (Persson et al. 
1980) and {\it HST} (Brown et al. 1998; King et al. 1995). 
The radio flux seems to have slight time variations. 
Among the {\it HST} data, the resolution of the $1.7\times10^{15}$ Hz 
observation is much better ($\sim 0^{\prime\prime}.5$, King et al. 1995) 
than the other two (at 1.1 and 1.5 $\times10^{15}$ Hz, Brown et al. 1998). 

Although the available data points are rather few, the Compton fitting
is very tight. 
Namely, if we try to adjust the fitting parameters such that the 
predicted curve passes through the radio and X-ray points (satisfying
both luminosity and spectral index), it does automatically also the 
$1.7\times10^{15}$ Hz point (Figure 4). 
Other {\it HST} and IR data should then be regarded as an excess 
whose origin may be attributed to other components than the ADAF 
under consideration. 
The fitted slope in the {\it Chandra} band seems somewhat harder, but 
it is the best we can do. 

The determined fitting parameters are cited in Table 1, and the
physical quantities of our interest are calculated as 
\begin{eqnarray}
  M = 7.0\times10^5\ M_{\odot}, \quad 
  \dot{M}_0 = 5.6\times10^{-7}\ M_{\odot}\ \mbox{yr}^{-1}, \nonumber\\
  r_{\rm out} = 3.7\times10^{14}\ \mbox{cm}, \quad 
  r_{\rm in} = 2.8\times10^{12}\ \mbox{cm} 
    = 1.4\times10^1\ r_{\rm S}, \nonumber\\
  \vert B_0\vert = 4.7\times10^{-1}\ \mbox{G}, \quad 
  \vert b_{\varphi}(r_{\rm in})\vert = 7.0\times10^{2}\ \mbox{G}, \nonumber\\
  \Delta = 8.73\times10^{-2}\ \mbox{rad}. 
\end{eqnarray}

On the other hand, if we adopt the bremsung fitting, the fitting 
to the X-ray slope can be improved much more. 
In Figure 5, there are three curves that demonstrate the variations 
caused by the change in the wind parameter. 
Other parameters are adjusted in each curve to fit the radio, UV and 
X-ray luminosities simultaneously. 
As predicted in the previous section, we can clearly see the main 
effects of winds on the spectra, i.e., the steepening of the bremsung 
fall toward the high frequency side and the suppression of the
synchrotron peak. 
For a more complete demonstration of such effects as caused by winds, 
we show in Figure 6 the variation of the spectrum according to the 
values of the wind parameter in its full range. 

The values of the fitting parameters for each curve are summarized
also in Table 1. 
The fitting to the X-ray slope seems satisfactory when $n$ is in 
between 0.3 and 0.5. 
However, it should be noted that the limiting case of $n=0.5$ 
corresponds to a flat density profile (i.e., it is independent of $x$; 
see Eq. [\ref{eqn:dens}]), which seems rather unplausible. 
The mass accretion rate in this fitting case increases by about one
order of magnitude compared with Compton fitting case. 

The resulting dimensional quantities in the case of thin solid curve
($n=0.3$), for example, are 
\begin{eqnarray}
  M = 4.0\times10^5\ M_{\odot}, \quad 
  \dot{M}_0 = 2.7\times10^{-6}\ M_{\odot}\ \mbox{yr}^{-1}, \nonumber\\
  r_{\rm out} = 5.3\times10^{16}\ \mbox{cm}, \quad 
  r_{\rm in} = 8.3\times10^{12}\ \mbox{cm} = 7.0\times10^1\ r_{\rm S}, 
\nonumber\\
  \vert B_0\vert = 6.0\times10^{-3}\ \mbox{G}, \quad 
  \vert b_{\varphi}(r_{\rm in})\vert = 2.2\times10^{2}\ \mbox{G}, 
\nonumber\\
  \Delta = 9.02\times10^{-4}\ \mbox{rad}. 
\end{eqnarray}

If we accept the bremsung fittings, we can conclude that M\ 31 has a 
considerably strong wind ($0.3<n<0.5$).
However, it is again difficult to clearly insist the superiority of
the bremsung fitting to the Compton fitting because of the shortage
of the observational data points. 
High resolution observations at mm-waves are desirable for this purpose. 

\subsection{Discussion}

In the previous subsections, we could reproduce the observed broadband
spectra of Sgr A$^*$ and M\ 31 fairly satisfactorily, by both of the
Compton and bremsung fittings.
Although some superiority of the latter case may be recognized in
the fitting to the X-ray slope, it is difficult to say clearly which
is the better one from the goodness of the fitting only.
However, the preference becomes much more clear when we combine
other information than from their spectra.
In this course, the main issue relating to our basic model will also
become evident. 

The central dark mass of Sgr A$^*$ is measured by the method of tracing 
stellar trajectories, and the result is $2.61\pm0.35\times10^6\ 
M_{\odot}$ within the inner 0.015 pc (Eckert \& Genzel 1997; Ghez 
et al. 1998). 
On the other hand, the point mass obtained from our spectral fittings 
is almost one order of magnitude smaller than the `dynamical' mass. 
For M\ 31, the central mass is estimated to be $3.3\times10^7\ M_{\odot}$
by using {\it HST} photometry (Kormendy \& Bender 1999). 
Also in this case, our model largely under-estimates it (apparently by 
about 30$\sim$40 times), even if the ambiguity caused by the shortage 
of data points is taken into account. 

Since in our model the black hole mass is determined essentially 
by the fitting to the luminosity of self-absorbed part of the
synchrotron emission, the above fact means that our model is 
over-estimating this luminosity.
This can be confirmed by the dashed curve in Figure 7 which is
written by adopting the dynamical mass ($m=2.6\times10^{-2}$) without 
changing other parameters from the case of solid curve in Figure 3.
The main reason for this over estimation may be attributed to
the over estimation of electron temperature near the disk's inner 
edge, in spite of the resulting rather large inner-edge radii
(a few tens of $r_{\rm S}$) than in other models. 

This over estimation is avoided in the viscous ADAF models by
introducing a two temperature scheme in which the electron temperature
deviates from a virial-type ion temperature, especially in an inner
region ($x<100$), owing to the increasing synchrotron loss and
inefficient energy supply from ions (e.g., Narayan \& Yi 1995b;
Narayan 2002).
This may indeed be a plausible explanation, but it is still uncertain 
whether or not the effective collisions between ions and electrons
can remain sufficiently small even in the presence of strong turbulence
in the accretion flows (Balbus \& Hawley 1998; Kaburaki, Yamazaki 
\& Okuyama 2002). 

In addition to the increase of radiation cooling in the inner
most accretion disks, there may be other boundary effects
such as launching of a jet, which are not included in our idealized
treatment of the accretion flow under the assumption of large
magnetic Reynolds number. 
Another possibility may be the effect of viewing angle of the
self-absorbed part of the disk, though it has not been considered
in the current ADAF fittings except Manmoto (2000). 
Indeed, we see only a smaller effective area compared with the
face-on looking, when the viewing angle becomes the closer to the
edge-on.
However, this effect can scarcely amount to more than an order
of magnitude.

As for the mass accretion rate, it is widely believed that the
Bondi accretion rate gives a plausible estimate of the actual
accretion rate at the remote (i.e., at the Bondi radius) regions
(e.g., Melia \& Falcke 2001).  
However, we have to emphasize here that it is nothing more than
a {\it fiducial} value like the Eddington accretion rate.
This is because there is no definite evidence for Bondi's spherical
accretion processes to be realized generally in actual places.
Instead, the validity of that model has been criticized explicitly 
by Narayan (2002). 

Anyway, {\it Chandra} provides a good estimate of the Bondi
accretion rate for the case of Sgr A$^*$.
Baganoff et al. (2001a) reports that there is $\sim$ 2 keV gas
with the number density $n_0\sim 100$ cm$^{-3}$ spread over about
an arcsec around the central source. This size is consistent with
the Bondi radius, $r_{\rm B}=2GM/V_{\rm B}^2$ where $V_{\rm B}$
is the sound velocity at $r_{\rm B}$, and the resulting Bondi
accretion rate is $\dot{M}_{\rm B} \equiv 4\pi r_{\rm B}^2
\rho_{\rm B}V_{\rm B} \sim 3\times10^{20}$ gs$^{-1}$
$\sim 5\times10^{-6} M_{\odot}$yr$^{-1}$.

Since this value corresponds to a normalized accretion rate of
the order of $\dot{m}_{\rm B}\sim10^{-4}$ for the observed dynamical
mass, we show in Figure 7 the prediction of our model in this case 
by fixing the other parameters also at the same values as in
the case of thick solid curve. 
It is evident that the result over estimates the observed luminosity
by about three orders of magnitude.
We can recognize also in this figure that the overall luminosity
is much more sensitive to the mass accretion rate than to the
central mass.
Therefore, uncertainties in the mass determination causes not so
drastic effects on the determination of mass accretion rate and
other parameters. 

Our best fit value for the normalized mass accretion rate is
$\sim0.04\times \dot{m}_{\rm B}$ and the affairs are similar also
for the resistive ADAF fittings (Narayan 2002).
Since the outer edge of the ADAF in our bremsung fitting case
(and also in the fittings by Narayan 2002) is of the order of
the Bondi radius, the large discrepancy between the predicted mass
accretion rates and the Bondi rate suggests that the accretion,
at least, in Sgr A$^*$ does not proceed like Bondi's prediction.
Although there is no firm reasoning to relate our $\dot{M}_0$
to the Bondi rate $\dot{M}_{\rm B}$, a simple relation
$\dot{M}_0/\dot{M}_{\rm B}\sim\Delta$ seems to account for the results
of spectral fittings fairly well.
The ratio represents the geometrical fraction of a disk accretion with
a half-opening angle $\Delta$ compared with the spherical accretion. 
This relation should be compared with a similar one in the viscous
ADAF models, $\dot{M}_{\rm N}/\dot{M}_{\rm B}\sim\alpha$ where
$\alpha$ is the viscosity parameter (Narayan 2002).

As already been pointed out in KKY, the low-frequency radio ($\nu<86$ 
GHz) excess seen above our fitting curves to Sgr A$^*$ is very likely
to come from the inner jet that is located within the inner edge of
the accretion disk and extend to the vertical directions along the
polar axis. 
Although its existence in the Galactic center, in particular, has not 
been established observationally yet, the universality of the disk-jet 
association seems plausible on both theoretical (see \S 2 of this paper) 
and observational (Ho 1999; Nagar, Wilson, \& Falcke 2001; Ulvestad 
\& Ho 2001) grounds. 
There are already many such models in which the spectrum of Sgr A$^*$ 
is reproduced by the jet-only or jet-plus-disk model (e.g., Falcke \& 
Markoff 2000; Yuan 2000; for a more comprehensive review, see Melia \& 
Falcke 2001).
Therefore, we have to check the above idea by including the 
contribution from a jet in our calculation of spectra. 
However, this is beyond the scope of the present paper.

The major concern with the case of the Compton fittings would be the 
resulting large external magnetic fields. 
The typical value obtained for it is $\sim$1 G, and this value 
is uncomfortably large as the ambient values near the outer edge,
even if a pre-amplification of the interstellar magnetic field caused
by the sweeping effect of accreting flow is taken into account. 
On the other hand, this value decreases to $\sim$ a few
$\times 10^{-3}$ G in the bremsung case.

The bremsung fittings predict that a widely spreading ADAF exists
with a very small half-opening angle of $\Delta\sim10^{-3}$
-$10^{-2}$, in each of the objects under our consideration.
It is very favorable to this result that {\it Chandra} seems to
have resolved the diameter of the central source in Sgr A$^*$
as 1 arcsec.
This corresponds to a radius of $6\times10^{16}$ cm and is in very
good agreement with our result of the bremsung fitting.
Combined with the consideration in the previous paragraph, we can 
say that we have fairly good reasons for the preference of the
bremsung fittings to the Compton fittings. 

If such a geometrically thin disk as predicted by our fittings is
surrounded by a tenuous wind-plasma, the configuration is somewhat
reminiscent of the disk-corona structure in the evaporation model
(e.g., Meyer, Liu, \& Meyer-Hoffmeister 2001).
However, the equatorial disk in our case is an ADAF, not an optically 
thick disk of standard type. 
The greatest concern with this situation would be the global stability 
of such a widely-spreading twisted magnetic structure. 
At present, this is an open question. 
In this connection, we only quote a recent work by Tomimatsu, Matsuoka, 
\& Takahashi (2001) in which a stabilizing effect of rotating magnetic 
fields on the screw instability is reported.

\section{SUMMARY AND CONCLUSION}

We have examined the expected effects of a wind on the emerging 
spectrum from an ADAF in a global magnetic field, based on the 
recently proposed resistive ADAF model including winds (K01). 
The main effects are seen both in the spectral index (the power of $\nu$) 
appearing in the intermediate frequency range of the thermal bremsstrahlung, 
and in the luminosity ratio of the thermal synchrotron emission to 
the bremsstrahlung. 
These two values decrease according to the strength of a wind. 
This fact can be explained by a suppressed mass accretion rate in the 
inner disk caused by wind loss. 

In order to test the plausibility of the resistive ADAF model, 
we have fitted the observed broadband spectra of Sgr A$^*$ and 
of the nucleus of M\ 31, by this model. 
For each observed spectrum, there are two possible types of fittings. 
One is the Compton fitting in which the negative X-ray slopes in 
the spectral energy distribution, which are obtained by {\it Chandra} 
for both objects, are reproduced by the synchrotron self-Compton process,
and the other is the bremsung fitting in which the negative slopes 
are reproduced by the intermediate frequency range of the thermal 
bremsstrahlung. 

On the grounds of the goodness only of the fitting to the
observational data points currently available, it is difficult to
clearly distinguish the superiority of the one type of fitting to
the other, because of the shortage of the observational data. 
However, we prefer the bremsung fittings for both objects. 
The main reasons for that are uncomfortably large values required 
for the strength of the seed magnetic field in the Compton fittings 
(0.5-3 G for both objects) and the very wide extension of accretion 
disks in the bremsung fittings, which seems favorable to the X-ray 
observations in the case of Sgr A$^*$. 

If the bremsung fittings are more plausible, we can conclude that 
the ADAFs extend so far as to reach the Bondi accretion radii 
(for both objects $r > (1$ - $5)\times10^5\ r_{\rm S}$), with no or 
very weak wind in the Sgr A$^*$ case, and with fairly strong wind 
in the M\ 31 case.
The resulting mass accretion rates for both objects are smaller
than the Bondi rates by more than an order of magnitude, and this
fact strongly suggests that the actual accretion processes in these
objects are certainly different from Bondi's spherical accretion. 

The major concern of our model in its present form is in the point
that it largely under estimates the central masses. 
This fact seems to come from the over estimation of the electron
temperature near the disk's inner edge.
Therefore, the improvement of the model's accuracy especially
near the inner boundary is strongly desired. 
 
\par

\section*{Acknowledgments}

We are grateful to Michael Garcia for correspondence about their data, 
and to Makoto Hattori for helpful discussions on X-ray observations. 
We also appreciate the valuable comments by the anonymous referee,
which have deepened our understanding on the role of the Bondi
accretion rate. 



\begin{figure}
\includegraphics[width=8cm]{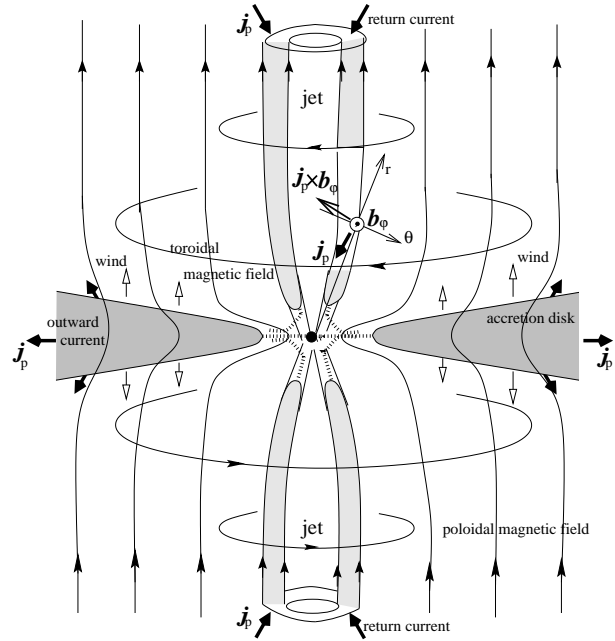}
\caption
%
{Conceptual drawing of the global configuration. 
The poloidal and toroidal components of the magnetic field lines 
are drawn with thin solid lines. 
The wind emanating from the disk surfaces and the poloidally 
circulating electric current are shown with white and black arrows, 
respectively.
Also shown is the Lorentz force acting on the return current in 
the polar region, which has the components both for collimation 
and for radial acceleration of a jet.
We distinguish jets from winds by their location and by the 
mechanisms for their launching. }
\end{figure}

\begin{figure}
\includegraphics[width=8cm]{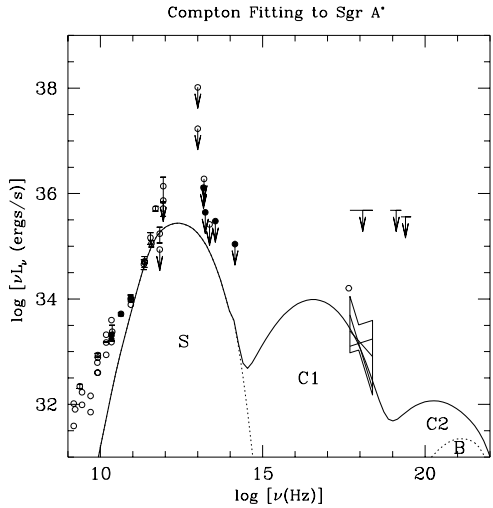}
\caption
%
{Compton fitting to Sgr A$^*$. The {\it Chandra} X-ray spectrum 
(Baganoff et al. 2001a) is fitted by the shoulder of the once 
Compton-scattered peak (C1) of the synchrotron photons (S). 
The thin dotted curve represents the corresponding spectrum 
in which the Compton-scattered components are suppressed. 
It can be seen that the contribution from bremsstrahlung (B) is buried 
in C2.
The open circle just above the {\it Chandra} error box indicates 
the luminosity observed by {\it ROSAT}. 
Other observational data points are compiled by Narayan et al. 
(1998b). 
In this type of fittings, the wind is suppressed ($n=0$). }
\end{figure}

\begin{figure}
\includegraphics[width=8cm]{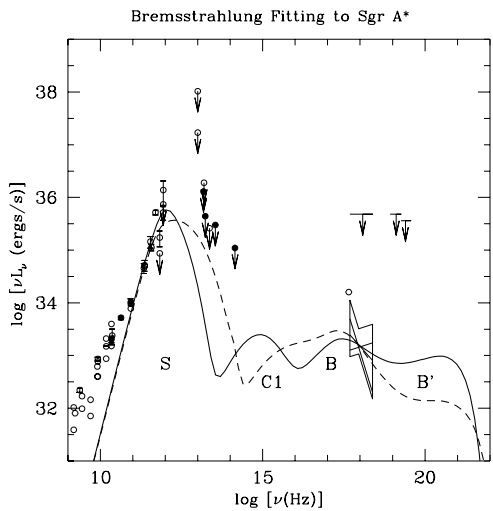}
\caption
%
{Bremsstrahlung fittings to Sgr A$^*$. 
The {\it Chandra} X-ray spectrum is fitted by negative slopes 
appearing in the intermediate frequency range of thermal 
bremsstrahlung spectra. 
The solid curve represents the no wind case ($n=0$), while the 
dashed curve does a case of weak wind ($n=0.1$). 
As seen in each curve, the contribution from the bremsstrahlung 
has double peaks.
The higher frequency peak (B$^{\prime}$) is due to an enhanced Gaunt
factor in the relativistic temperature range. }
\end{figure}

\begin{figure}
\includegraphics[width=8cm]{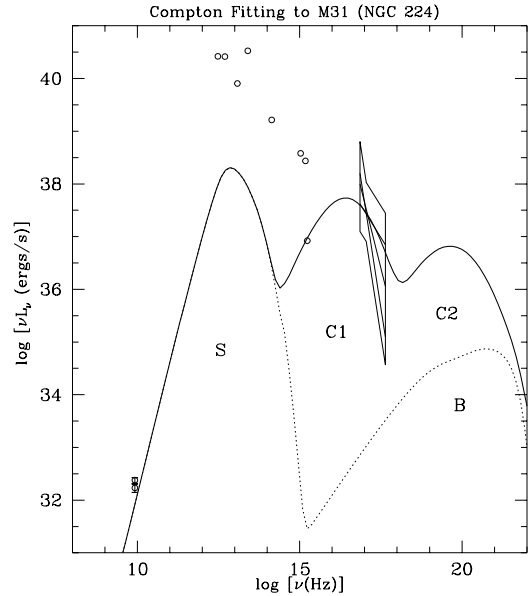}
\caption
%
{Compton fitting to M\ 31. 
The solid curve is the almost uniquely determined best fit curve, 
with no wind ($n=0$). 
The thin dotted curve represents the corresponding spectrum 
in which the Compton-scattered components are suppressed. 
Also in this case, it can be seen that the contribution from 
bremsstrahlung (B) is buried in C2.}
\end{figure}

\begin{figure}
\includegraphics[width=8cm]{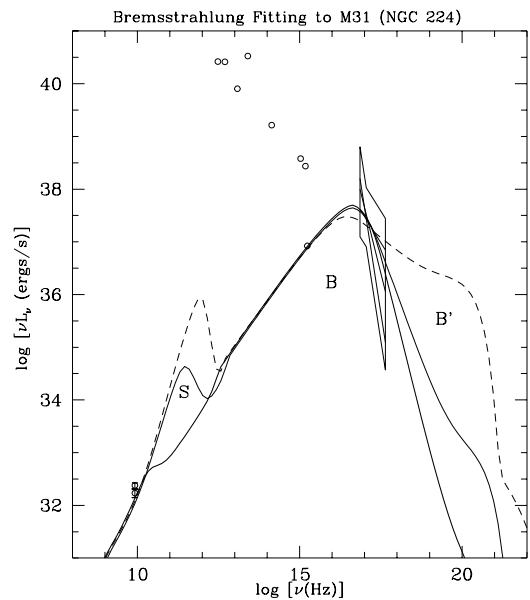}
\caption
%
{Bremsstrahlung fittings to M\ 31. 
The dashed curve is the best fit curve with no wind ($n=0$). 
It is clear that the negative slope in the {\it Chandra} band 
cannot be reproduced without winds. 
However, both the high frequency shoulder of the bremsstrahlung 
and the height of the synchrotron peak become suppressed with 
the increasing wind strength: $n=0.3$ for the thin solid curve, 
and $n=0.5$ for the thick solid curve. 
Thus, the {\it Chandra} spectrum is well reproduced by these strong 
wind cases.}
\end{figure}

\begin{figure}
\includegraphics[width=8cm]{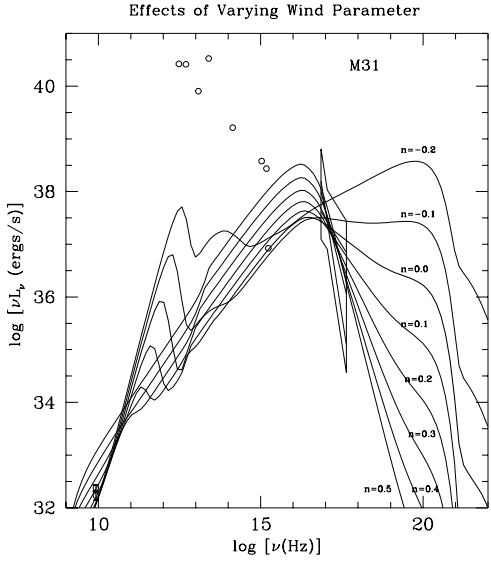}
\caption
%
{Demonstration of the wind effects. 
The wind parameter is varied almost in its full range, as indicated 
on each curve.
The base is the dashed curve in figure 5 ($n=0$), and the other 
parameters than $n$ are all fixed. 
The negative values of $n$ mean the downward winds toward the disk 
surfaces. }
\end{figure}

\begin{figure}
\includegraphics[width=8cm]{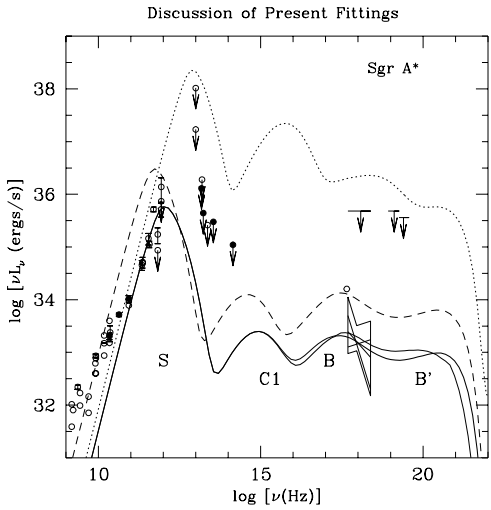}
\caption
%
{Comparison of various fittings. As a reference curve, the best fit
curve with no wind in Fig. 3 (thick solid curve) is adopted. The thin
solid curve is the prediction by our old version of the calculation
scheme, for the same values of the fitting parameters. Their
difference results from the difference in the adopted Gaunt factors.
The thin curve runs slightly above the thick curve until it falls
rapidly at higher frequencies. The small dip above the peak B is
artificial. The dashed and dotted curves are the predictions in the
present scheme, when the central mass alone is replaced by the
dynamical mass and when the accretion rate alone is replaced by the
Bondi rate, respectively.}
\end{figure}

\onecolumn

\begin{table}
\caption{Summary of model parameters}
\begin{tabular}{cccccccc}
\hline
Figure & Fit. type & Line type & $n$ & $m$ & $\dot{m}$ & $x_{\rm out}$ & $R_{0}$ \\
\hline
2 & C & solid & $0.0$ & $6.0\times10^{-3}$ & $7.2\times10^{-7}$ & $1.0\times10^{2}$ & $3.84$ \\
3 & B & solid & $0.0$ & $4.0\times10^{-3}$ & $4.1\times10^{-6}$ & $1.0\times10^{5}$ & $5.88\times10^{1}$ \\
  &   & dashed & $0.1$ & $4.0\times10^{-3}$ & $2.3\times10^{-6}$ & $1.0\times10^{5}$ & $8.70\times10^{1}$ \\
4 & C & solid & $0.0$ & $7.0\times10^{-3}$ & $3.6\times10^{-5}$ & $1.8\times10^{3}$ & $1.15\times10^{1}$ \\
5 & B & dashed &  $0.0$ & $3.0\times10^{-3}$ & $6.4\times10^{-4}$ & $1.0\times10^{6}$ & $8.00\times10^{1}$ \\
  &   & thin solid &  $0.3$ & $4.0\times10^{-3}$ & $3.0\times10^{-4}$ & $4.5\times10^{5}$ & $8.00\times10^{1}$ \\
  &   & thick solid &  $0.5$ & $1.0\times10^{-2}$ & $1.1\times10^{-4}$ & $4.0\times10^{5}$ & $8.00\times10^{1}$ \\
6 & B & thin solid & $-0.2$ & $3.0\times10^{-3}$ & $6.4\times10^{-4}$ & $1.0\times10^{6}$ & $8.00\times10^{1}$ \\
  &   & thin solid & $-0.1$ & $3.0\times10^{-3}$ & $6.4\times10^{-4}$ & $1.0\times10^{6}$ & $8.00\times10^{1}$ \\
  &   & thick solid & $0.0$ & $3.0\times10^{-3}$ & $6.4\times10^{-4}$ & $1.0\times10^{6}$ & $8.00\times10^{1}$ \\
  &   & thin solid & $0.1$ & $3.0\times10^{-3}$ & $6.4\times10^{-4}$ & $1.0\times10^{6}$ & $8.00\times10^{1}$ \\
  &   & thin solid & $0.2$ & $3.0\times10^{-3}$ & $6.4\times10^{-4}$ & $1.0\times10^{6}$ & $8.00\times10^{1}$ \\
  &   & thin solid & $0.3$ & $3.0\times10^{-3}$ & $6.4\times10^{-4}$ & $1.0\times10^{6}$ & $8.00\times10^{1}$ \\
  &   & thin solid & $0.4$ & $3.0\times10^{-3}$ & $6.4\times10^{-4}$ & $1.0\times10^{6}$ & $8.00\times10^{1}$ \\
  &   & thin solid & $0.5$ & $3.0\times10^{-3}$ & $6.4\times10^{-4}$ & $1.0\times10^{6}$ & $8.00\times10^{1}$ \\
7 & B & thick solid & $0.0$ & $4.0\times10^{-3}$ & $4.1\times10^{-6}$ & $1.0\times10^{5}$ & $5.88\times10^{1}$ \\
  & B & dashed & $0.0$ & $2.6\times10^{-2}$ & $4.1\times10^{-6}$ & $1.0\times10^{5}$ & $5.88\times10^{1}$ \\
  & B & dotted & $0.0$ & $4.0\times10^{-3}$ & $1.0\times10^{-4}$ & $1.0\times10^{5}$ & $5.88\times10^{1}$ \\
  & B$^{*}$ & thin solid & $0.0$ & $4.0\times10^{-3}$ & $4.1\times10^{-6}$ & $1.0\times10^{5}$ & $5.88\times10^{1}$ \\
\hline
\end{tabular} \\
Notes: C;Compton fitting, B;Bremsstrahlung fitting,
B$^{*}$;Bremsstrahlung fitting (old version)
\end{table}

\twocolumn

\end{document}